\documentclass[12pt,fleqn]{article}

\usepackage{epsfig}
\usepackage{amsmath}
\usepackage{times}
\usepackage[sort&compress,comma]{natbib}
\usepackage[hang,small]{caption}

\newcommand{\Eq}[1]{Equation~(\ref{-1})}
\newcommand{\eq}[1]{Eq.~(\ref{-1})}
\newcommand{\eqs}[1]{Eqs.~(\ref{-1})}
\newcommand{\fig}[1]{Fig.~\ref{-1}}
\newcommand{\Fig}[1]{Figure~\ref{-1}}
\newcommand{\note}[1]{\marginpar{\bf -1}}

\makeatletter
\renewcommand\@biblabel[1]{#1.}
\makeatother

\begin{document}
\title{\sffamily\textbf{Speciation due to hybrid necrosis in plant-pathogen
    models}}
\vspace{0.5cm}
\author{Iaroslav Ispolatov \& Michael Doebeli\\\
\vspace{-2mm}\normalsize Department of Zoology and Department of Mathematics, \\
\vspace{7mm}\normalsize University of British Columbia, Vancouver B.C., Canada
V6T 1Z4} 
\date{\vspace{10mm}\normalsize\today}
\maketitle
\vskip 10mm
\noindent 
\vskip 10 mm
\noindent {\bf Corresponding author:} Michael Doebeli, doebeli@zoology.ubc.ca
\vskip 1 cm
\noindent {\bf Keywords}: speciation, evolution of diversity, immune system, autoimmune reaction, hybrid necrosis, hybrids

\vskip 10 mm

\def\d{\delta}
\def\D{\Delta}
\def\s{\sigma}
\def\g{\gamma}
\def\e{\epsilon}
\def\b{\beta}
\def\a{\alpha}
\def\l{\lambda}
\def\L{\Lambda}
\def\k{\Kappa}

\maketitle

\newpage
\noindent{\Large \bf Abstract}
\vspace{5mm}

We develop a model for speciation due to postzygotic
incompatibility generated by autoimmune reactions. The model is based on predator-prey interactions between a
host plants and their pathogens. Such interactions are often frequency-dependent, so that pathogen attack is focused on the most abundant plant phenotype, while rare plant types may escape pathogen attack. Thus, frequency dependence can generate disruptive selection, which can give rise to speciation if distant phenotypes become reproductively isolated. Based on recent experimental evidence from {\it Arabidopsis}, we assume that at the molecular level, incompatibility between strains is caused by epistatic interactions between two proteins in the plant immune system, the guard and the guardee. Within each plant strain, immune reactions occur when the guardee protein is modified by a pathogen effector, and the guard subsequently binds to the guardee, thus precipitating an immune response. However, when guard and guardee proteins come from phenotypically distant parents, a hybrid's immune system can be triggered by erroneous interactions between these proteins even in the absence of pathogen attack, leading to severe autoimmune reactions in hybrids. Our model shows how phenotypic variation generated by frequency-dependent host-pathogen interactions can lead to postzygotic incompatibility between extremal types, and hence to speciation.

\maketitle

\section{Introduction}
Understanding the origins of diversity is a central theme in evolutionary
biology. In general, evolutionary diversification can be described as temporal
modification of phenotype distributions (\cite{coyne_orr2004, doebeli_etal2007}). A single species typically
corresponds to a unimodal phenotype distribution, whereas this distribution
becomes bimodal (or multimodal) once diversification has occurred (\cite{doebeli_etal2007}). Thus,
speciation can be described as a splitting of the ancestral unimodal phenotype
distribution into a two or more descendant peaks, with each peak corresponding
to an emerging species. Traditionally, diversification and speciation are
thought to occur when different phenotypes are selectively favoured in
different and isolated geographical regions, so that an initially uniform
ancestral population that is geographically dispersed would develop different
phenotypic modes corresponding to the phenotypes favoured in different
locations. More recently, processes of adaptive speciation, which unfold in
the absence of geographical isolation and during which phenotype distributions
become multimodal due to ecological interactions such as competition for
resources of predation, have received considerable attention (\cite{dieckmann_etal2004}). In sexual
populations, adaptive speciation requires that the different clusters in the
phenotype distribution, corresponding to the newly emerging species, are
separated by barriers to gene flow. That is, to prevent mixing  between the
emerging species, successful reproduction should occur predominantly within each
emerging phenotypic cluster, and not between individuals with distinctly
different phenotypes. The type of reproductive isolation that has been
considered in most theoretical models of adaptive speciation is isolation due
to assortative mating, i.e., prezygotic isolation
(\cite{dieckmann_doebeli1999}, \cite{dieckmann_etal2004}). In this paper, we investigate the potential role of postzygotic isolation for adaptive speciation by considering models in which postzygotic isolation is caused by autoimmune responses.

Recently, \cite{eizaguirre_etal2009} have pointed out that the immune system may play an important role in processes of diversification. These authors primarily considered the role of the vertebrate immune system for prezygotic isolation (so that mating partners would be chosen based on immune system characteristics), but they also mentioned that the immune system may be important for postzygotic isolation due to a weakened immune response in hybrids. Here we consider a different type of hybrid disadvantage due to malfunctioning of the immune systems, which is is based on observations made in plant systems. In plants, a well-known example of postzygotic isolation is
hybrid necrosis, defined as a set of highly deleterious and often lethal
phenotypic characteristics (\cite{bomblies_weigel2007}). In hybrid necrosis, a mixture of genes from
different strains becomes deleterious even though the contributing genes were
harmless, or even beneficial, in the parents. Recent experimental evidence suggests that hybrid necrosis in 
plants can be caused by an epistatic interaction of loci
controlling the immune response to attack by pathogens (\cite{bomblies_etal2007}). In this form of hybrid necrosis, inviability is caused by inappropriate activation of the plant
immune system in the absence of pathogens. Among several mechanisms of pathogen recognition by a plant host cell,
interactions between two different types of host proteins, ``guard'' and
``guardee'' proteins, are thought to play a key role 
(\cite{jones_dangl2006}).  When a pathogen attacks a host cell,  it often injects
effector proteins that manipulate target proteins in the host cell and thereby
contribute to the success of the pathogen. These host targets (the guardees)
are ``guarded'' by other host proteins (the ``guard'') that monitor the
guardee's molecular structure. When a pathogen effector induces changes in the
molecular structure of the guardee (creating ``pathogen-induced modified-self''), these
changes are recognized by the guard proteins, which then activate the immune
response (\cite{jones_dangl2006}). Thus, on the one hand successful pathogen attack
requires fine-tuning of the effector to the guardee, so that a pathogen with a
given effector repertoire can only successfully attack a certain range of host
cells (i.e., those with the ``right'' types of guardees). On the other hand,
efficient immune response requires fine-tuning of the guardee and the guard,
so that the guard selectively recognizes only those guardees that have been modified by a
pathogen. However, if mating between different host strains leads
to hybrids in which guard and guardees come from lineages with different
evolutionary paths, the guardee might recognize the guard as modified even in
the absence of pathogen attack, which could lead to immune response and
subsequent necrosis of the hybrid even in the absence of any
pathogen. Experimental evidence of two-locus epistatic interactions for hybrid
necrosis, of the autoimmune nature of the deleterious phenotype, and of an
increased disease resistance of rescued from necrosis hybrids  all support the hypothesis
that guard-guardee interactions are involved in hybrid necrosis in the well-studied model plant {\it
  Arabidopsis thaliana} (\cite{bomblies_etal2007}). A large number of hybrid necrosis cases sharing phenotypic
similarities with the {\it Arabidopsis} cases indicate that this may 
be a common mechanism operating in a wide range of plant species.

Due to the
potentially high selection pressures exerted by pathogen attack, genes
controlling immune responses are generally thought to be fast evolving. This can in turn generate strong selection pressures on pathogens, which can lead to coevolution. In addition, such host-pathogen, or more generally, predator-prey interactions, are often frequency-dependent, and it is known that this frequency-dependence can generate disruptive selection in the host (\cite{doebeli_dieckmann2000, dercole_etal2003}). Here we present a mathematical model of adaptive
speciation in a host plant in which diversification is by driven host-parasite
interactions, and postzygotic reproductive isolation is caused by hybrid
necrosis. The barriers to gene flow between
emerging phenotypes are generated by detrimental autoimmune
reactions in individuals containing pathogen-resistance genes from different clusters. 
Specifically, reproductive postzygotic isolation is due to genetic incompatibility between guard and guardee proteins of
phenotypically distant strains. In the following we develop a simple representation of the evolution of the
guard and guardee proteins driven by selection for escaping recognition of the
guardee by pathogen effectors. As we will show, the frequency-dependent
selection imposed by the pathogen leads to the emergence of distinct strains
that are reproductively isolated due to genetic incompatibility in the immune
response. The emerging host strains correspond to two distinct paths of
coordinated evolution of guard and guardee proteins. Mating between these
strains can result either in hybrids that are very disease-prone due to a
defective immune system, or in individuals that show hybrid necrosis due to
autoimmune reactions.

\section{Model description}
\subsection{Phenotype space}

Several experimental observations from \cite{bomblies_etal2007} are essential for the
definition of our model. First, it has been uncovered that the autoimmune 
reaction 
plays an essential role in lethality of the hybrid phenotype, and that the
surviving hybrids exhibit increased pathogen resistance. Second, it was shown 
that epistatic interactions between two loci are both a necessary and a
sufficient condition for hybrid necrosis. And finally, it was observed  
that an increase in a habitat temperature from $16^{\circ}$ to $23^{\circ}$
rescues the hybrid. All these phenomena strongly indicate the possibility that
hybrid necrosis is 
caused by erroneous binding between guard and guardee proteins coming
from different parents.  In normal plants, such binding only occurs if the
guarded protein is modified by the pathogen effector. However, in a hybrid such
binding could happen in the absence of pathogen effectors since the guard and
guardee proteins evolved independently and could thus have acquired a
propensity for binding, i.e., a higher binding energy,  
without any additional
alterations through pathogen effectors. An increase in the ambient temperature
weakens any binding, thus decreasing a chance to provoke the unwarranted
immune response in a hybrid, exactly as observed in \cite{bomblies_etal2007}. An estimate
showing that the increase in ambient temperature applied in the experiments
can indeed cause a noticeable shift in binding equilibrium of guard and
guardee proteins is presented in Appendix A. For our model, we envisage
biochemical binding between guard and guardee proteins to be the mechanism for
both immune and autoimmune  responses. 

We assume that the evolution of guardee and guard protein is described by two
phenotypic coordinates, $g$ and $r$. Each host plant individual is represented
by a point in this two-dimensional phenotype space, with the $g$-coordinate
describing the state or genetic makeup of the individual's guardee protein, and
the $r$-coordinate describing the state or genetic makeup of the individual's
guard protein. Mutations in guard and guardee genes cause the corresponding 
point to shift in $(r,g)$ space. The density of plant individuals with a particular
form of guard and guardee proteins is described by a density distribution function $h(r,g)$. An almost homogeneous population with little genetic
variation in immune proteins is described by a density distribution with a single narrow peak, while a
population consisting of several strains with different guard and guardee
proteins is described by a multimodal density distribution.  

In our framework, the pathogen is characterized by a single coordinate, $e$,
which describes the genetic makeup of the parasite's effector proteins. The
effector proteins interacts with the guardee protein of a host plant,
and we define $e$, as in more traditional predator-prey models (e.g. \cite{doebeli_dieckmann2000}), so that
the pathogen attack is most effective for small distances $\vert e-g\vert$
between the pathogen phenotype $e$ and the host's guardee phenotype $g$. Thus,
it is in the interest of the host to have a guardee phenotype that is
different from the phenotype of the most common pathogen, and this  is  the
mechanism that generates frequency-dependent selection.  

In the plant host, the relative values of the $r$ and $g$ coordinates
determine the immune reaction 
of the plant to a parasite effector, as well as autoimmune reactions. We
assume that there is a trade-off between mounting an efficient immune response
in the presence of pathogen effectors, and being prone to deleterious
autoimmune reactions in the absence of pathogens. Specifically, we assume that
when  $g >> r$, the probability that the guard binds to the guardee protein
and triggers an immune response is small, independent of whether pathogen
effectors are present or not.  In this case, the plant immune system is less
sensitive, making the plant more susceptible to pathogen attack, but less
prone to autoimmune reactions. Conversely, when $r > >g$, the  guard protein
has a high propensity to bind to the guardee. In this case, the plant immune system is
more sensitive, making the plant less susceptible to pathogen attack, but more
prone to autoimmune reactions. This parametrization of phenotype space is in
an accord with the  observation made in \cite{bomblies_etal2007} that necrotic hybrids,
being rescued by 
elevated ambient temperature, are very effective in suppressing pathogen
attack.  

Quantitatively, we assume that the probability for an $(r,g)$-plant to die
from parasitic infection once attacked by a pathogen is proportional to
$\exp[(g-r)/\sigma_I]$, while the probability to die from autoimmune reactions is
proportional to $\exp[(r-g)/\sigma_{AI}]$, where $\sigma_I$ and $\sigma_{AI}$ are system parameters. As a consequence of this trade-off, plants
tend to survive best if their phenotypes satisfy $r\approx g$ case. 
\subsection{Plant and parasite evolution}
 
To describe the co-evolution of the plant and pathogen populations, we use a
predator-prey-style model for the dynamics of phenotype distributions in both
the host plant and the pathogen, with specific terms that describe the  
plant-parasite conflict and autoimmune reactions. Mathematically, the model is
a system of two integro-differential equations that gives the temporal
evolution of the host density distribution $h(r,g)$ and the pathogen density
distribution $p(e)$. In the integrals, the quantity $h(r,g) dr dg$ is the density of plants
with guard and guardee phenotypes in the coordinate interval $(r, r+dr)$ and $(g,
g+dg)$, and the quantity $p(e) de$ is the density of pathogens with effector
phenotype in the coordinate interval $(e, e+de)$. The equations determining the
rate of change in the distributions $h(r,g)$ and $p(e)$ consist of birth and death terms,
describing the increase and decrease in plant and parasite population
densities due to the various biological components and interactions, as
follows.

\begin{itemize}
\item For a given plant phenotype $(r,g)$, the rate of attack from the
  pathogen, and hence the probability of infection, is proportional to a
  weighted sum over all pathogen phenotypes, with the weights reflecting how
  well a pathogen phenotype $e$ can attack a plant with guardee phenotype
  $g$. The weight function, or ``attack kernel'', is assumed to be of Gaussian
  form and given by  
\begin{align}
\label{attack}
G_a(e-g)=\exp\left[-\frac{(e-g)^2}{2\sigma_a^2}\right],
\end{align}
reflecting the fact that infection is easiest if the pathogen phenotype is
very similar to the guardee phenotype (on some appropriate scale).  
Accordingly, for the given plant phenotype $(r,g)$, the probability of being
attacked is proportional to  
\begin{align}
\int p(e)  G_a(e-g) de,
\end{align}
where $p(e)$ is the pathogen density distribution. Once attacked, the
probability that the plant individual dies is determined by the immune
response (as described above) and is  proportional to $\exp[(g-r)/\sigma_I]$,
reflecting how well the guard protein recognizes the changes in the guardee
protein induced by the pathogen attack. Overall, this leads to a total death
rate of plants with phenotypes $(r,g)$ due to pathogen attack given by 
\begin{align}
\label{dh}
 - \d_I h(r,g) \exp\left(\frac{g-r}{\sigma_I}\right) \int p(e)  G_a(e-g) de.
\end{align}
Here $\d_I$ is a constant of proportionality, and $h(r,g)$ is the density of
host plants with phenotype $(r,g)$. The minus sign indicates that this is a
death term.

\item In the plant population, death also occurs due to autoimmune reactions,
  whose magnitude in plants of phenotype $(r,g)$ is proportional to
  $\exp[(r-g)/\sigma_{AI}]$. The resulting death rate is given by 
\begin{align}
\label{dha}
- \d_I h(r,g) \exp\left(\frac{r-g}{\sigma_{AI}}\right).
\end{align} 
For simplicity, we assume that the rate coefficient $\d_I$ for the autoimmune
and parasite-induced death terms is the same.

\item To complete the death terms  for the plant, we assume that density
  dependent competition in the absence of the pathogen results in a logistic
  death term of the form 
\begin{align}
\label{dch}
- \d_C \frac{h(r,g) H(t)}{K(r,g)}.
\end{align}
here $H(t)$ is the total density of the plant population, i.e.
\begin{align}
H(t)=\int_{r,g}h(r,g)drdg.
\end{align}
The parameter
$\d_C$ in eq. (5) is a rate coefficient for the plant death rate due to
competition. The function $K(r,g)$ is the carrying capacity function, which we assume to be
of the form  
\begin{align}
K(r,g)=\exp\left[-\frac{(r-r_0)^2}{2\sigma_r^2}\right]\exp\left[-\frac
{(g-g_0)^2}{2\sigma_g^2}\right]. 
\end{align}
This reflects the assumption that due to costs and benefits of the guard and
guardee that are unrelated to their effect on immune response, there are
optimal values $r_0$ and $g_0$ for these traits. This means that in the
absence of pathogen attack and immune response, the plant traits would evolve
to their optimal values $r_0$ and $g_0$ (and the plant distribution would
converge to a narrow unimodal distribution centred at $(r_0,g_0)$). In this way,
the carrying capacity function $K$ provides a component of stabilizing
selection that is independent of host-pathogen interactions. 

In general, it is not clear what the position of $(r_0,g_0)$ would be in phenotype space. In particular, it is not clear what cost stabilizing selection would impose on a well-tuned immune response, for which $r \approx g$. Therefore, we assume in the following that only the trait associated with the guardee ($g$) protein affects the carrying capacity. As this is also the trait that affects the host-pathogen interaction, this corresponds to commonly made assumptions (e.g. \cite{doebeli_dieckmann2000}). Mathematically, that fact the the $r$-trait does not affect the carrying capacity corresponds to the assumption that the width of the carrying capacity in the $r$-direction is very large, $\sigma_r \rightarrow \infty$.

\item
To derive the birth term for the plant population, we need to incorporate
sexual reproduction, and for simplicity we assume that individuals are haploid
and have two loci with continuously varying alleles encoding the two
phenotypes $r$ and $g$. An offspring with phenotype $(r',g')$ either inherits
the two alleles $r'$ and $g'$ from different parents, or from the same
parent. In the first case, the offspring comes from a mating between
$(r',g'')$ and $(r'',g')$ (or vice versa) with all possible $g''$ and $r''$. Such matings occur with probability proportional to
$\frac{ h(r',g'') h(r'',g')}{2H(t)}$, where $H(t)$ is the total plant density
as before. In the second case, the probability that such an offspring is
produced is simply proportional to $h(r',g')$. Adding the two cases together, the total
probability that an $(r',g')$ offspring is the result of mating is thus 
\begin{align}
 \int \frac{ h(r',g'') h(r'', g')}{2H(t)} dr''dg'' + \frac{h(r',g')}{2}
\end{align} 
(the factor $1/2$ reflects ambiguity in assigning mother and father in any
given mating pair). 
In addition, we assume that mutation in the plant traits is described by two normal mutation
kernels 
\begin{align}
G_{M,r}(r-r')
=\frac{1}{\sqrt{2\pi}\sigma_{M,r}}\exp\left[-\frac{(r-r')^2}
{2\sigma_{M,r}^2}\right] 
\end{align}
and
\begin{align}
G_{M,g}(g-g')
=\frac{1}{\sqrt{2\pi}\sigma_{M,g}}\exp\left[-\frac{(g-g')^2}{2\sigma_{M,g}^2}
\right].  
\end{align}
Here $G_{M,r}(r-r')$ describes the probability that a mutation in an
$r'$-allele produces a value in the interval $[r,r+dr]$, and similarly for
$G_{M,g}(g-g')$. Thus, small mutations are more likely than large ones. 

Overall, the rate at which offspring with phenotype $(r,g)$ are produced is
then given by 
\begin{align}
\label{bh}
\b \int \int  G_{M,r}(r-r') G_{M,g}(g-g') \left[ \int \int \frac{ h(r',g'') 
h(r'',g')}{2H(t)} dr''dg'' + \frac{h(r',g')}{2}\right] dr' dg',
\end{align} 
where $\b$ is the birth rate. 

\item For the pathogen, birth is determined by successful infection. For
  pathogen type $e'$, the probability that it successfully attacks and
  subsequently infects a host of type $(r,g)$ is proportional to  
\begin{align}
h(r,g) \exp\left(\frac{g-r}{\sigma_I}\right) G_a(g-e'), 
\end{align}
where $G_a$ is the attack kernel given by eq. (1). Therefore, the total probability for
pathogen type $e'$ to successfully infect any host plant is  
\begin{align}
\int \int h(r,g)
  \exp\left(\frac{g-r}{\sigma_I}\right) G_a(g-e') dr dg.
\end{align}
For simplicity, we assume that reproduction is asexual in the pathogen, but as
in the plant species we include mutation given by a normal function $G_{M,e}$
(with a width described by a parameter $\sigma_{M,e}$, so that the total rate
of production of offspring of type $e$ is 
\begin{align}
\label{bp}
\a \d_I \int G_{M,e}(e-e') p(e') \left[ \int \int h(r,g)
  \exp\left(\frac{g-r}{\sigma_I}\right) G_a(g-e') dr dg \right ] de'.
\end{align}
Here the
conversion coefficient $\a$ indicates how many new parasites are produced from 
an infected host.  

\item Finally, as in many other predator-prey models the death rate of the
  pathogen is assumed to be 
\begin{align}
\label{dp}
 - \d_P p(e),
\end{align}
for some parameter $\d_P$ describing the intrinsic per capita death rate of the pathogen. Note that we assume that this death rate is independent of the
pathogen phenotype $e$.  
\end{itemize} 
   
\vskip 1 cm

\noindent
Collecting all the birth and death terms into two equations describing the
dynamics of the plant and pathogen density distributions now yields the
following system of coupled partial differential equations: 
\begin{align}
\label{prp}
\nonumber
\frac {\partial h(r,g)} {\partial t} = 
&\b \int \int  G_{M,r}(r-r') G_{M,g}(g-g') \left[ \int \int \frac{ h(r',g'') 
h(r'',g')}{2H(t)} dr''
  dg''  + \frac{h(r',g')}{2}\right] dr' dg'\\
&- \d_I h(r,g) \exp\left(\frac{g-r}{\sigma_I}\right) \int p(e) G_a(e-g) de\\
\nonumber
&- \d_I h(r,g) \exp\left(\frac{r-g}{\sigma_I}\right)
- \d_C \frac{h(r,g) H(t)}{K(r,g)}, \\
\nonumber
\\
\nonumber
\frac {\partial p(e)} {\partial t} = 
&\a \d_I \int G_{M,e}(e-e')p(e') \left[ \int \int h(r,g) \exp\left(\frac{g-r}{\sigma_I}\right)
G_a(g-e') dr dg\right] de'\\
& - \d_P p(e). 
\end{align}

These nonlinear equations have many parameters and in principle may exhibit a
variety of dynamic regimes. Due to the apparent complexity of the dynamical system, we do not expect any analytical results to be feasible and
instead investigated this system using numerical simulations. We are
particularly interested in those regimes that lead to multimodal equilibrium
distributions in the host plant.

 \begin{figure}[htp]
\includegraphics[width=.45\textwidth]{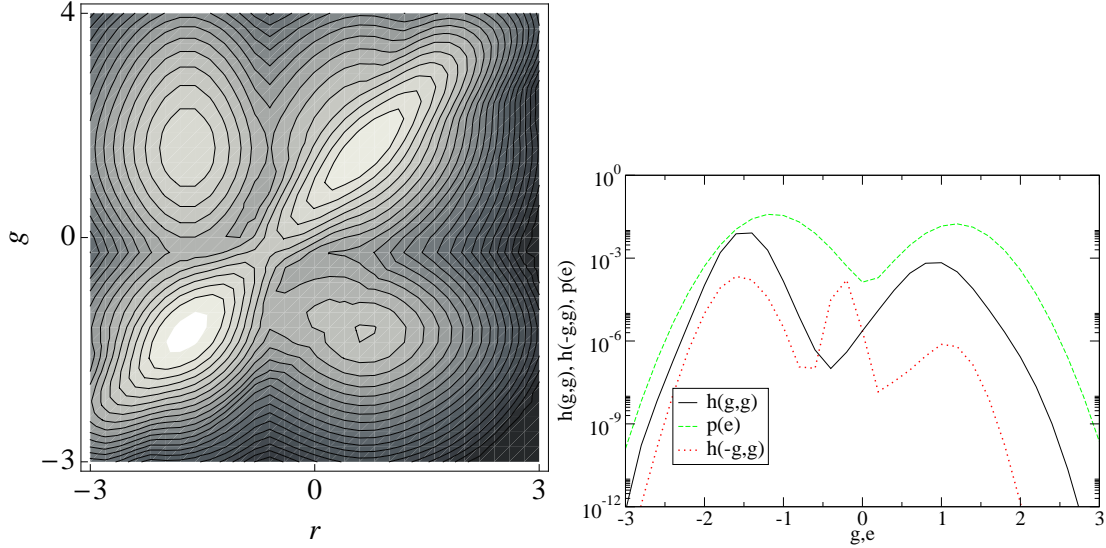}
\includegraphics[width=.45\textwidth]{fig1a.eps}
\caption{\label{Fig. 1}
Left panel: Contour plot of the host population density in logarithmic scale. 
Clearly visible along the $r=g$ diagonal are 
two peaks, centered around two coordinates $(r_1,g_1)$ and $(r_2,g_2)$,
corresponding to two strains, driven into separation by parasite-plant
interactions. Two much weaker peaks with coordinates approximately
$(r_1,g_2)$ and 
$(r_2,g_1)$, correspond to hybrid heterozygotes in the $R$ and $G$
genes. The subpopulation of necrotic hybrids lies below the $r=g$
diagonal, where $r >g$. The plot corresponds to the solution of 
eqs. (16) and (17) at a steady state ($t=800$) for the following parameter values:
$\b=1$, $\a=1$, $\d_I=5$, $\d_C=1$, $\d_P=1$, $\sigma_I=0.2$, $\sigma_{AI}=0.2$, $\sigma_a=0.8$,
$\sigma_{M,r}=\sigma_{M,g}=\sigma_{M,e}=0.1$, and $\sigma_r=\sigma_g=1$. To avoid unrealistically high autoimmune and pathogen-induced host death rates, they were truncated by replacing $\exp \pm\left(\frac{r-g}{\sigma}\right)$
by $\min \left[ \exp \pm\left(\frac{r-g}{\sigma}\right), 10^3\right]$.
Right panel: Plot of the host population $h(r,r)$ along the $r=g$ diagonal (black solid line) and
the parasite 
population $p(e)$ (red dashed line). 
}
\end{figure}

 Such
dynamics  are illustrated in Figure 1 and occur for a substantial range of parameters and correspond to cases
in which the host-pathogen interaction leads to plant speciation.
Intuitively, the choice of parameters generating this scenario can be explained as follows: The width of the carrying capacity $\sigma_g$ 
defines the phenotypic space scale of the model and is taken to be one. One of the birth coefficients, $\b_h$, defines the time scale and is also taken to be one. Finally, the competition death term defines the third scale, the amplitude of the population density and is taken to be one as well. The mutation width for the host $\sigma{M,r}$, 
$\sigma_{M,g}$
and pathogen $\sigma_{M,e}$ define the minimal width of the emerging pattern and should be significantly smaller than one. To enable pattern formation due to host-parasite interaction, the parasite attack width should also be less than one. Finally, the intensity of the autoimmune reaction and the susceptibility to the parasite infection should grow sufficiently fast (faster than carrying capacity decay to be relevant) as the phenotypic distance from the optimum, $r\approx g$, increases. Thus, the immune and autoimmune reaction widths $\sigma_I$ and $\sigma_{AI}$ must be noticeably less than one as well. Other rates are set equal to unity except for the coefficients for the autoimmune and parasite-induced death term, $\d_I$, which, to make this term more significant, is set equal to 5.  

In Figure 1,
each of the main maxima in the host plant distribution, located at
$(r_1,g_1)$ and $(r_2,g_2)$, respectively, corresponds to an emerging host
strain or sub-species with a distinct genetic makeup of guard and guardee
proteins. Note that $r_1\approx g_1$ and $r_2 \approx g_2$, which means that
in each of the corresponding strains, the function of the guard and guardee
proteins are geared towards both efficient immune response to the pathogen and
low likelihood of autoimmune reactions. The equilibrium distribution in Figure
1 also shows two secondary peaks located approximately at $(r_1,g_2)$ and
$(r_2,g_1)$. These peaks correspond to hybrids between the two emerging
subspecies. These hybrids inherit their guard and guardee genes from parents
of different subspecies and thus are either easy victims of pathogen attack
(when $g>r$) or exhibit strong autoimmune necrosis (when $r>g$). Despite the
equal per capita rate of birth of the homozygote and heterozygote offspring (see 
eq. (11)), the peaks corresponding to the hybrids are much smaller due
to their much higher death rates. This is illustrated in Figure 1b, which shows the bimodal equilibrium density distributions of the host (continuous line) and pathogen (dashed line) along the diagonal, as well as the trimodal density distribution of the host along the anti-diagonal (stippled line). Here the mode in the middle corresponds to the saddle between the two modes of the host distribution along the diagonal.

Obviously, the model can also produce other equilibrium distributions, depending on the values of model parameters, particularly the parasite attack width. Examples are given in Figures 2 and 3. In Figure 2, the distributions converge to a unimodal equilibrium. In accordance with previous results (\cite{doebeli_dieckmann2000, dercole_etal2003}), this tends to happen for larger predator attack widths $\sigma_a$. In contrast, for small attack widths, the equilibrium distributions may have more than two modes along the diagonal, and hence more than two secondary hybrid peaks, as illustrated in Figure 3.

\begin{figure}[htp]
\includegraphics[width=.45\textwidth]{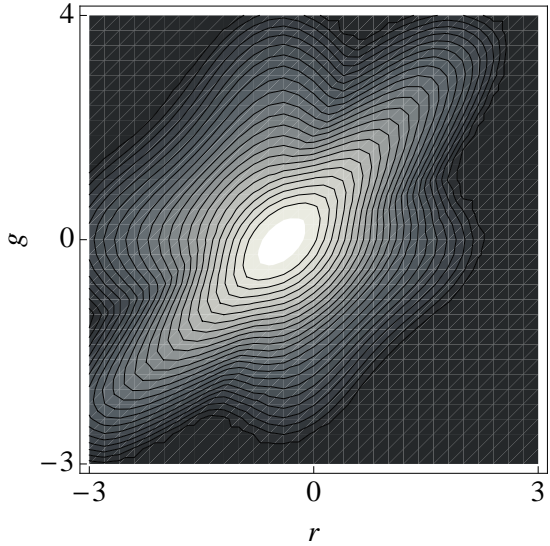}
\caption{\label{Fig. 2}
Same as Fig. 1, but for a larger parasite attack width, $\sigma_a=1.2$.
In this case the host density does not evolve to a multimodal distribution, but forms a broad single peak, and the same is true for the pathogen density distribution (not shown).} 
\end{figure}

\begin{figure}[htp]
\includegraphics[width=.45\textwidth]{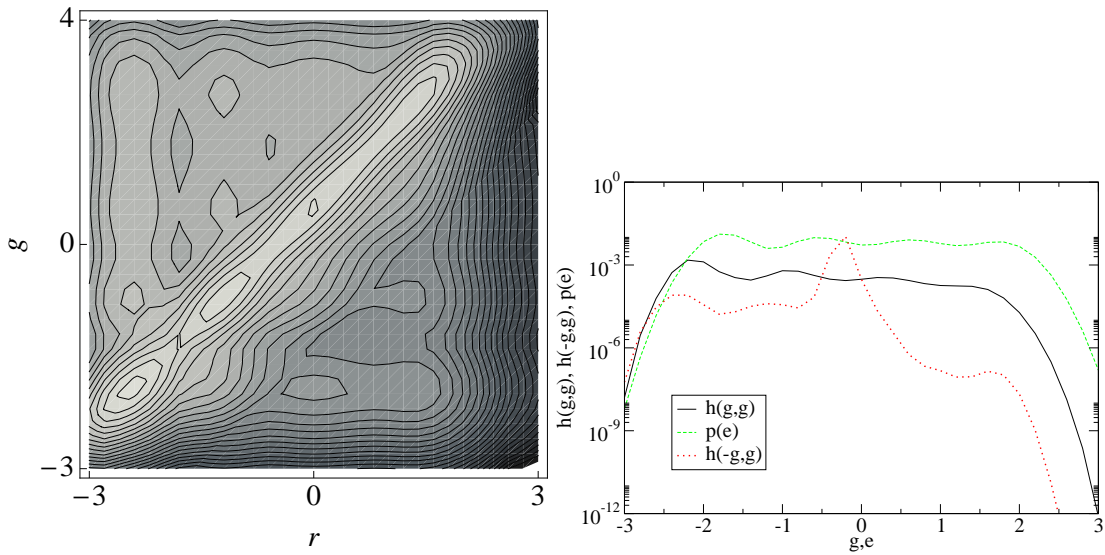}
\includegraphics[width=.45\textwidth]{fig3a.eps}
\caption{\label{Fig. 3}
Same as Fig. 1, but for a smaller parasite attack width $\sigma_a=0.4$.
In this case the host density evolves to a multimodal distribution, with multiple secondary peaks corresponding to hybrids (Fig. 2a). The multiple peaks are also evident if the host distribution is restricted to the diagonal, and the pathogen distribution also has multiple peaks (Fig. 2b). Along the anti-diagonal, the host density distribution reflects the multiple secondary hybrid peaks (Fig. 2b).} 
\end{figure}

%


\section{Discussion}
Both \cite{bomblies_weigel2007} and \cite{eizaguirre_etal2009} recently argued that the immune system could play an important role in speciation processes. While \cite{eizaguirre_etal2009} mainly considered the vertebrate immune system as a potential source of prezygotic reproductive isolation after divergent adaptation to pathogens, \cite{bomblies_weigel2007} argued that defective autoimmune reactions in hybrids could be the source of postzygotic isolation. They were able to support this perspective by impressive experimental evidence in {\it Arabidopsis thaliana} (\cite{bomblies_etal2007}). Our aim here was to provide mathematical evidence for the feasibility of adaptive speciation due to host-pathogen interactions when isolation is caused by postzygotic autoimmune deficiencies.

\cite{bomblies_etal2007} have shown that hybrid necrosis, defined as a set of deleterious and often lethal
phenotypic characteristics, can be caused by an epistatic interaction of loci
controlling the immune response to attack by pathogens in {\it Arabidopsis thaliana}. More precisely, hybrid necrosis is caused by detrimental activation of the plant
immune system in the absence of pathogens, and \cite{bomblies_etal2007} showed that epistatic interactions between two loci are both necessary and sufficient for this form of hybrid necrosis. This corresponds to classic Dobzhansky-Muller incompatibilities (\cite{gavrilets2004}), which we have modeled here by assuming that the optimal compromise between the ability to respond to pathogen attack and the avoidance of undesired autoimmune reactions is attained by individuals with genotypes $(r,g\approx r)$, where $r$ and $g$ represent two loci controlling the immune response. Thus, two genotypes $(r,r)$ and $(R,R)$ with $R$ very different from $r$ can both be optimal, but their heterozygous hybrids $(r,R)$ and $(R,r)$ will suffer from a malfunctioning immune system. In the case of the plant immune system, the two genes $r$ and $g$ represent two different proteins, ``guard'' and
``guardee'' proteins. Such proteins are thought to be central for the plant immune system (\cite{jones_dangl2006}), and their optimal functioning requires concerted evolution at both loci.

The model presented here combines a detailed, albeit schematic, description of the plant immune response based on the traits $r$ and $g$, with the macroscopic, population-based representation of ecological interactions driving the evolution of these traits. The basic result is that for a range of intuitively appealing parameters, the model leads to adaptive speciation: as a result of frequency-dependent selection exerted by pathogen attack, a single ancestral host strain splits into two descending strains $(r,r)$ and $(R,R)$ that each have their distinct genetic makeup of the guard-guardee pathogen recognition system. Hybrids with guard and guardee genes coming from different parents either suffer necrosis due to autoimmune reaction, or they exhibit a weakened immune system due to a compromised ability to detect the parasite attack. This latter form of hybrid disadvantage corresponds to the mechanism of postzygotic isolation due to MHC-based immune responses that was conjectured by \cite{eizaguirre_etal2009} to operate in vertebrates.

In contrast to most previous models of adaptive speciation, in which reproductive isolation is based on prezygotic mechanisms such as assortative mating, in our models reproductive isolation between diverging lineages emerges due to postzygotic hybrid inviability. Of course, a natural question would be whether such postzygotic isolation would select for prezygotic isolation in the form of assortative mating based on the immune system, as envisaged by \cite{eizaguirre_etal2009}. Our models could be extended to pursue this, as well as a number of other extensions. For example, it is known that many pathogens are capable of producing a number of different effector proteins that allow them to attack a host plant. Similarly, the immune system of the plant has a number of different ways of dealing with this effector variety. Accordingly, it would be interesting to see whether the type of diversification observed in our models would be easier or harder to obtain if the dimensionality of both the guard and guardee traits in the host and the effector trait in the pathogen is increased. For example, it is possible that hybrid necrosis, and hence postzygotic isolation, is increased if more than one guard-guardee pairs are involved in the immune response to a particular pathogen, because simultaneous incompatibility of various guards and guardees could lead to more severe autoimmune reactions. 
                                                                
Instead of considering larger numbers of guard-guardee pairs, another way to increase the the complexity of the model would to develop a more detailed, mechanistic description of the genetic network of activation and repression of the various pathways involved in the immune response based on a single guard-guardee pair. In the present model, this network is assumed to be extremely simple in that it is assumed that the binding affinity of guard to guardee depends on the genetic distance $(r-g)$. \cite{tentusscher_hogeweg2009} have studied more traditional models for adaptive speciation based on resource competition under the assumption that the regulatory network determining the phenotypes important for competition are much more complicated, and realistic, than commonly assumed in such models. One of the main conclusion of these authors is that genetic complexity facilitates diversification and speciation, and it would be interesting to see whether similar conclusions would be reached if more genetic complexity would be incorporated into the models presented here.

Despite being schematic and minimalistic, our model correctly reflects some of the main experimental observations of postzygotic isolation due to hybrid necrosis in {\it Arabidopsis thaliana} (\cite{bomblies_etal2007}, see also Appendix). It also predicts a class of hybrids with weakened immune response to pathogen attack, whose existence should be investigated in future experiments. Overall, we agree with \cite{eizaguirre_etal2009} that studying the role of the immune system for speciation processes is very promising. Adaptation in the immune system as a co-evolutionary response to pathogens may be a potent mechanism for diversification based on both prezygotic and postzygotic reproductive isolation.

\section{Appendix}
Here we provide an argument for the fact that an increase in ambient temperature from $16^o$ to $23^o$ can indeed cause a biologically noticeable reduction in autoimmune reaction and hence rescue otherwise necrotic hybrids, as observed in the experiments of \cite{bomblies_etal2007}. We make a simple estimate of how a change in temperature affects the Law of
Mass Action governing the binding-unbinding equilibrium between the guard and
guardee proteins. For simplicity we assume that both protein can exist only in
two forms, free and bound to each other forming a dimer. we denote the free
forms by $R$ and $G$, and the dimer by $RG$. In the limit of weak binding
(large dissociation constant $k$), i.e., when both proteins are mostly 
in free form, the concentration of the dimer is       
\begin{align}
\label{lma}
  [RG]=[R]_0 [G]_0 / k,\tag{A1}
\end{align}
where $[R]_0$ and $[G]_0$ are the total concentrations of proteins $R$ and
$G$. The temperature dependence of the dissociation constant is usually
given by the Arrhenius form, 
\begin{align}
\label{ar}
  k=k_0 \exp \left(- \frac{E}{KT}\right).\tag{A2}
\end{align}
As a consequence, the relative decrease in the concentration of $RG$ caused by
the increase in 
temperature by $\Delta T$ is
\begin{align}
\label{drg}
  \frac{[RG]_{T+\Delta T}} {[RG]_T}  = 
\exp \left(- \frac{E \Delta T}{KT^2}\right).\tag{A3}
\end{align}
A cell can typically recognize a change in concentration larger than 20\%
(smaller shifts in concentrations are 
apparently perceived as ``noise'', \cite{weissman_etal2006}). Thus, for a very reasonable
value of protein-protein dissociation energy $E\approx 8kT$, 
an increase in the temperature from   $16^{\circ}$ to $23^{\circ}$  can
produce a biologically meaningful decrease in concentration of the $RG$
dimer. It is at least plausible that such a decrease can eliminate improper
binding between $R$ and $G$, and hence unwanted autoimmune reactions.  

\newpage
\bibliography{arabidopsis}
\bibliographystyle{prslb}



\end{document}